%% file: main.tex
\documentclass{article}
\usepackage{spconf,amsmath,amssymb,graphicx}
\usepackage{bm}
\usepackage{booktabs,tabularx,multirow}
\usepackage{tikz}
\usepackage{fontawesome5}
\usepackage{xcolor}
\usetikzlibrary{arrows.meta,positioning,calc,fit,backgrounds}
\usepackage{microtype}
\usepackage{paralist}
\setdefaultleftmargin{1.2em}{}{}{}{}{}
\usepackage[hidelinks]{hyperref}
\usepackage{url}
\usepackage{orcidlink}

\newcommand{\snow}{\textcolor{cyan!55!black}{\faSnowflake}}
\newcommand{\fire}{\textcolor{orange!80!black}{\faFire}}

\let\oldthebibliography\thebibliography
\renewcommand{\thebibliography}[1]{%
  \oldthebibliography{#1}%
  \setlength{\itemsep}{-0.91pt}%
  \setlength{\parsep}{0pt}%
}

\title{CROSS-MODAL KNOWLEDGE DISTILLATION FOR ACOUSTIC PEDESTRIAN DETECTION}

\name{\shortstack{Yonghyun Kim$^{1}$\,\orcidlink{0000-0001-7541-9939}, \; Chaeyeon Han$^{2}$\,\orcidlink{0000-0001-7311-3944}, \; Sancho Gatungay$^{3}$\,\orcidlink{0009-0003-9706-7585}, \\[2pt]
      Subhrajit Guhathakurta$^{2}$\,\orcidlink{0000-0002-2456-3284}, \; Alexander Lerch$^{1}$\,\orcidlink{0000-0001-6319-578X}}}
\address{%
  $^{1}$Music Informatics Group, Georgia Institute of Technology, USA \\
  $^{2}$Center for Urban Resilience and Analytics, Georgia Institute of Technology, USA \\
  $^{3}$College of Computing, Georgia Institute of Technology, USA \\}

\begin{document}
\ninept
\maketitle

\input{sections/0_abstract}
\input{sections/1_introduction}

\input{sections/2_related}

\input{sections/3_method}

\input{sections/4_experiments}
\input{sections/5_results}

\input{sections/7_conclusion}

\bibliographystyle{IEEEbib}
\bibliography{refs}

\end{document}

%% file: sections/0_abstract.tex
\begin{abstract}
Audio-only pedestrian detection is attractive for urban sensing but limited by weak acoustic cues. An appealing strategy is cross-modal knowledge distillation, in which a video teacher supervises the audio student during training so that the deployed model runs on audio alone. Under this task's severe class imbalance and wide video-audio modality gap, however, what such distillation contributes is unclear. We introduce \emph{Trust-Filtered Distillation (TFD)}, which selectively suppresses teacher supervision on pedestrian samples, and interpret its logit formulation as conditional label smoothing under a shared temperature. Across ten distillation configurations under five-fold cross-session validation on ASPED, most methods yield modest gains in macro accuracy accompanied by small changes in PR-AUC. The main effect is higher no-pedestrian accuracy at the cost of lower pedestrian recall. Adding TFD to logit distillation strengthens this trade-off without improving mean PR-AUC. These findings clarify the benefits and limitations of selective cross-modal supervision by distinguishing operating-point shifts from discrimination gains in imbalanced acoustic detection.

\end{abstract}


%% file: sections/1_introduction.tex
\vspace{-10pt}
\section{Introduction}
\label{sec:intro}

Reliable pedestrian detection supports urban safety, autonomous driving, and smart-city planning. Audio sensors complement cameras since they are inexpensive and remain effective where vision fails (e.g., darkness, fog, occlusion)~\cite{kim2025audio}. But the acoustic signal is harder to analyze. Pedestrian cues are weak (e.g., footsteps) and easily masked by urban noise, limiting audio-only detectors.

We investigate cross-modal knowledge distillation (KD)~\cite{hinton2015distilling} from video to audio, using visual supervision during training while retaining audio-only inference. KD has been applied to audio-visual~\cite{aytar2016soundnet, huo2024c2kd, sarkar2024xkd} and audio-to-audio~\cite{ding_audio_2023} tasks. In our setting, the teacher observes the scene directly, providing a strong discriminative signal.

This task presents two challenges for cross-modal distillation: severe class imbalance (e.g., $8.5\%$ pedestrian presence in ASPED~v.a~\cite{seshadri2024asped}) and a wide video-audio modality gap. We examine how distillation affects discrimination and operating points in this setting, and whether per-sample selective supervision changes these effects. Our contributions are:
\vspace{5pt}
\begin{compactenum}
\item \textbf{Distinguishing accuracy gains from discrimination gains.} Across ten KD configurations on ASPED, modest macro-accuracy gains often accompany lower pedestrian recall, with little improvement in distinguishing pedestrian presence from absence.

\item \textbf{Class-selective teacher supervision.} We introduce \emph{Trust-Filtered Distillation (TFD)}, a ground-truth-conditioned gate that suppresses minority-class KD, and show that it admits a conditional label-smoothing interpretation under a shared temperature.
\vspace{2pt}
\item \textbf{Separating probability calibration from detection ability.} We show that predicted pedestrian probabilities better match observed frequencies mainly through reduced overestimation, with little improvement in distinguishing pedestrian from non-pedestrian samples. Improved probability calibration therefore need not imply better detection in this setting.

\end{compactenum}

%% file: sections/2_related.tex
\vspace{-1pt}
\section{Related Work}
\label{sec:related}

\textbf{Acoustic pedestrian detection.}
Within acoustic urban sensing~\cite{bello2019sonyc, mesaros2019sed}, Seshadri et al.~\cite{seshadri2024asped} introduced audio-based pedestrian detection with synchronized audio, video, and pedestrian labels, and Kim et al.~\cite{kim2025audio} extended it to vehicular noise. Both report faint pedestrian sounds and labels heavily skewed toward no-pedestrian, countered with weighted sampling and loss, yet audio-only accuracy remains limited.

\vspace{3pt}
\textbf{Cross-modal KD.}
SoundNet~\cite{aytar2016soundnet} pioneered visual-to-audio transfer by training an audio network using visual-recognition outputs as supervision. Subsequent work explicitly addresses the modality gap. C2KD~\cite{huo2024c2kd} selects samples for distillation based on the rank agreement between teacher and student soft predictions, while XKD~\cite{sarkar2024xkd} introduces a domain-alignment objective between audio and video features. Nevertheless, successful transfer depends on information shared across modalities, not teacher accuracy alone~\cite{xue2023modality}. Modality difference can limit the benefits of distillation~\cite{chen2024nontarget}, and mismatched supervision may lead to negative transfer~\cite{chen2021distilling, wang2019characterizing}. Closer to detection, MM-DistillNet distills visual teachers into an audio student for sound-only object detection~\cite{valverde2021mmdistillnet}, and AV-PedAware distills a LiDAR teacher for audio-visual 3D pedestrian awareness~\cite{yang2023avpedaware}. We instead analyze video-to-audio distillation for audio-only pedestrian presence detection under severe class imbalance.

\vspace{3pt}
\textbf{KD variants and calibration.}
Beyond logit distillation~\cite{hinton2015distilling}, other variants match intermediate features (FitNets~\cite{romero2015fitnets}), use contrastive objectives (CRD~\cite{tian2020crd}), or preserve sample relations (RKD~\cite{park2019rkd}). Another line of work interprets logit distillation as regularization akin to label smoothing~\cite{yuan2020revisiting, szegedy2016rethinking}, with implications for predictive calibration~\cite{muller2019smoothing, guo2017calibration}. A recent study~\cite{kim2025teachercalib} reports that teacher calibration error correlates more strongly with student accuracy than teacher accuracy does for image classification on CIFAR-100.

\vspace{3pt}
\textbf{Class imbalance and selective KD.}
Common approaches include class-balanced re-weighting~\cite{cui2019classbalanced}, focal loss~\cite{lin2017focal}, and logit adjustment~\cite{menon2021logit}. Class imbalance has also been studied in KD for long-tailed recognition~\cite{bkd2023, deitlt2024}. Other methods select or reweight teacher supervision: FedLMD masks majority-class logit dimensions~\cite{fedlmd2023}, Hard Gate KD switches between teacher and ground truth by a calibration criterion~\cite{lee2022hardgate}, confidence-aware multi-teacher distillation weights teachers by ground-truth agreement~\cite{zhang2022camkd}, and Self-Filtered Distillation uses trust scores~\cite{sfd2025}. These studies leave open the systematic evaluation of selective distillation that combines sample-level selection, ground-truth guidance, and class-specific exemption in cross-modal detection under severe imbalance.

%% file: sections/3_method.tex
\vspace{-3pt}
\section{Method}
\label{sec:method}

This section defines the training losses and the proposed per-sample mask. Models, data, and training settings are given in Section~\ref{sec:experiments}.
\vspace{-4pt}
\subsection{KD Loss under Class Imbalance}
\label{sec:setup}

KD trains the student to match the teacher's softened predictions alongside the ground-truth labels. Let $\mathbf{z}_s^{(i)}, \mathbf{z}_t^{(i)} \in \mathbb{R}^{2}$ denote the student's and teacher's logits for training sample $i$, let $y_i \in \{0,1\}$ be its ground-truth label ($0$ for no-pedestrian, $1$ for pedestrian), let $\sigma(\cdot)$ be the softmax, whose component $\sigma(\cdot)_c$ is the probability of class $c \in \{0,1\}$, and let $T$ be the distillation temperature. Throughout, a capital $\mathcal{L}$ denotes a batch-level loss and a lower-case $\ell^{(i)}$ its per-sample term. Over a mini-batch of $N$ samples, the logit KD loss is
\begin{equation}
\mathcal{L}_{\mathrm{KD}} \;=\; \frac{T^{2}}{N}\sum_{i=1}^{N}\mathrm{KL}\!\left(\sigma(\mathbf{z}_t^{(i)}/T)\,\big\|\,\sigma(\mathbf{z}_s^{(i)}/T)\right),
\label{eq:kd_loss}
\end{equation}
and the total training loss is $\mathcal{L} = \mathcal{L}_{\mathrm{CE}} + \alpha\,\mathcal{L}_{\mathrm{KD}}$, where $\mathcal{L}_{\mathrm{CE}}$ is the cross-entropy (CE) against $y_i$ and $\alpha$ is the KD weight. Beyond this logit loss, we also evaluate a hybrid variant (Hybrid) that adds a cosine feature-alignment term between the teacher and student embeddings. The mask defined below applies to whichever KD term is used.

Prior work addresses class imbalance in distillation through class-aware objectives~\cite{bkd2023, deitlt2024}. In our setup the sampler balances the two classes in expectation (Section~\ref{sec:experiments}), removing the systematic overrepresentation of the majority class. The per-sample KD gradient drives the student's softened distribution toward the teacher's, $\partial\ell_{\mathrm{KD}}^{(i)}/\partial\mathbf{z}_s^{(i)}\propto\sigma(\mathbf{z}_s^{(i)}/T)-\sigma(\mathbf{z}_t^{(i)}/T)$, and so depends on both networks. One asymmetry we can observe directly is in the teacher target. With $p_t^{(i)}=\sigma(\mathbf{z}_t^{(i)})_1$ the teacher's pedestrian probability, its mean probability on the correct class (measured at $T{=}1$) is lower on pedestrian than on no-pedestrian samples ($\approx\!0.75$ vs.\ $\approx\!0.85$). We treat this asymmetry as an observed property of the supervision rather than a proven cause. We establish the net operating-point shift (Section~\ref{sec:opshift}) and its calibration signature (Section~\ref{sec:calibration}). Positive, student-independent weights rescale each sample's KD gradient, whereas a per-sample gate can remove selected KD contributions entirely.

\vspace{-4pt}
\subsection{Trust-Filtered Distillation}
\label{sec:tfd}

We propose \textbf{Trust-Filtered Distillation (TFD)}, in which a per-sample binary indicator $m^{(i)}\in\{0,1\}$, indicating whether KD is applied to sample $i$, gates the KD term,
\begin{equation}
\mathcal{L}_{\mathrm{KD}}^{\mathrm{TFD}} \;=\; \frac{\sum_{i=1}^{N} m^{(i)}\,\ell_{\mathrm{KD}}^{(i)}}{\sum_{i=1}^{N} m^{(i)}},
\label{eq:tfd}
\end{equation}
where $\ell_{\mathrm{KD}}^{(i)}$ is the per-sample KD term of Eq.~\ref{eq:kd_loss}. When $m^{(i)}{=}0$, sample $i$ contributes only its CE term. TFD masks the minority-class samples whose teacher target still carries substantial majority-class mass:
\begin{equation}
m^{(i)} \;=\; \begin{cases}
0 & \text{if}\ y_i=1\ \text{and}\ \sigma(\mathbf{z}_t^{(i)})_0 > \tau,\\
1 & \text{otherwise,}
\end{cases}
\label{eq:tfd_label}
\end{equation}
with $\tau{=}0.4$, the implementation default, not tuned on the folds. At $\tau{=}0.5$ the gate fires on minority samples the teacher misclassifies, and the lower value also silences those it gets right but with $\sigma(\mathbf{z}_t)_1{<}0.6$. The mask never touches majority-class samples. Since Eq.~\ref{eq:tfd} renormalizes by $\sum_i m^{(i)}$, masking also rescales the surviving KD weight by $N/\sum_i m^{(i)}$, and we set the masked KD loss to zero when no samples are retained.

\vspace{-4pt}
\subsection{TFD as Conditional Label Smoothing}
\label{sec:tfd_smoothing}

Building on the label-smoothing view of KD~\cite{yuan2020revisiting}, we read logit-based TFD, under a shared temperature, as conditional, teacher-dependent label smoothing. Masked samples retain the hard-label target, while unmasked samples use a mixture of the label and the teacher distribution. Formally, writing the per-sample objective in its unnormalized form $\ell_{\mathrm{CE}}^{(i)} + \alpha\,m^{(i)}\,\ell_{\mathrm{KD}}^{(i)}$:

\noindent\textit{Proposition.} Let $\mathbf{y}^{(i)}\in\{0,1\}^{2}$ be the one-hot label of sample $i$ and let $\ell_{\mathrm{CE}}^{(i)}$ be its ordinary (unweighted) cross-entropy term, with both losses taken at $T{=}1$. Up to an additive student-independent constant and a positive rescaling, the per-sample objective $\ell_{\mathrm{CE}}^{(i)} + \alpha\,m^{(i)}\,\ell_{\mathrm{KD}}^{(i)}$ equals the cross-entropy of the student's prediction against the smoothed target
\begin{equation}
\tilde{\mathbf{y}}^{(i)} \;=\; (1-\lambda^{(i)})\,\mathbf{y}^{(i)} + \lambda^{(i)}\,\sigma(\mathbf{z}_t^{(i)}),\qquad \lambda^{(i)} = \frac{\alpha\,m^{(i)}}{1+\alpha\,m^{(i)}},
\label{eq:tfd_smoothing}
\end{equation}
and for $T{>}1$ the same identity holds on the tempered teacher $\sigma(\mathbf{z}_t^{(i)}/T)$ and student $\sigma(\mathbf{z}_s^{(i)}/T)$ distributions when CE and KD share $T$, with $\lambda^{(i)} = \alpha T^{2} m^{(i)}/(1+\alpha T^{2} m^{(i)})$.

\noindent\textit{Sketch.} Expand $\ell_{\mathrm{KD}}^{(i)}$ into a teacher-weighted CE plus a student-independent constant, group the coefficients of the student's log-probabilities, and normalize their sum.

\noindent\textit{Scope.} The identity assumes ordinary, unweighted cross-entropy and a logit-KD term evaluated at the same temperature. Our experiments use a task loss at $T{=}1$ and a KD term at $T{=}3$, so Eq.~\ref{eq:tfd_smoothing} characterizes the shared-temperature case rather than exactly reproducing the experimental objective. For weighted cross-entropy at a shared temperature, the mixing coefficient also depends on the class weights and the loss normalization, including Eq.~\ref{eq:tfd}. The cosine term of the Hybrid variant is not covered by this identity.

\vspace{-4pt}
\subsection{Relation to Prior Selective KD}
\label{sec:tfd_related}

Prior methods select or weight supervision using calibration criteria~\cite{lee2022hardgate}, trust scores~\cite{sfd2025}, class-dimension masks~\cite{fedlmd2023}, or cross-modal prediction agreement~\cite{huo2024c2kd}. TFD instead combines a sample-level binary gate with an explicit ground-truth class condition. Eq.~\ref{eq:tfd_label} suppresses KD only for minority-class samples whose teacher assigns majority-class probability above the threshold. We compare how these methods select teacher supervision, without claiming that they are mathematically equivalent to Eq.~\ref{eq:tfd_smoothing}.

%% file: sections/4_experiments.tex
\vspace{-1pt}
\section{Experimental Setup}
\label{sec:experiments}

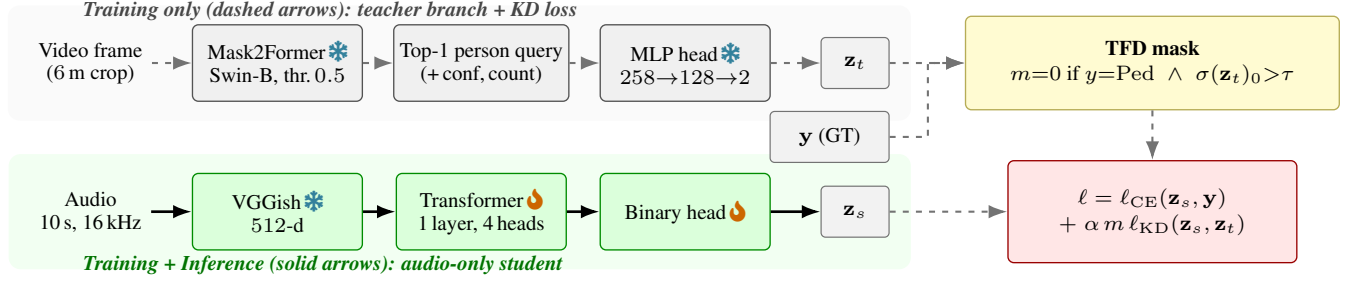
\begin{figure*}[t]
\centering
\resizebox{0.99\textwidth}{!}{%
\begin{tikzpicture}[
  font=\scriptsize,
  >={Latex[length=2.2mm,width=1.6mm]},
  arrsolid/.style={->, line width=0.9pt, draw=black},
  arrdash/.style={->, line width=0.7pt, draw=black!55, dash pattern=on 2pt off 2.4pt},
  baseblock/.style={draw, rounded corners=2pt, minimum height=9mm, minimum width=20mm, align=center, inner xsep=2pt, inner ysep=1.5pt, line width=0.45pt, font=\scriptsize},
  teacher/.style={baseblock, fill=gray!12, draw=gray!65!black},
  student/.style={baseblock, fill=green!14, draw=green!55!black},
  loss/.style={baseblock, fill=red!10, draw=red!60!black, minimum width=34mm, minimum height=12mm, font=\scriptsize},
  masknote/.style={baseblock, fill=yellow!22, draw=yellow!65!black, minimum width=44mm, minimum height=11mm, font=\scriptsize},
  logitnode/.style={draw=black!55, fill=gray!10, rounded corners=1.5pt, minimum height=6mm, minimum width=8mm, font=\scriptsize, inner sep=1.5pt, line width=0.4pt},
  inputlabel/.style={font=\scriptsize, draw=none, fill=none, align=center, minimum width=13mm},
  phaselab/.style={font=\scriptsize\bfseries\itshape},
]
\def\xA{0}\def\xB{2.2}\def\xC{4.6}\def\xD{7.0}\def\xE{9.0}
\def\xRight{12.5}
\def\xY{8.7}
\def\yT{0.75}\def\yS{-1.0}
\def\yY{-0.12}

\node[inputlabel] (vid)  at (\xA,\yT) {Video frame\\($6$\,m crop)};
\node[teacher]    (m2f)  at (\xB,\yT) {Mask2Former\,\snow\\Swin-B, thr.\,$0.5$};
\node[teacher]    (filt) at (\xC,\yT) {Top-1 person query\\(+\,conf,\,count)};
\node[teacher]    (mlp)  at (\xD,\yT) {MLP head\,\snow\\$258{\to}128{\to}2$};
\node[logitnode]  (zt)   at (\xE,\yT) {$\mathbf{z}_t$};
\foreach \a/\b in {vid/m2f, m2f/filt, filt/mlp, mlp/zt} \draw[arrdash] (\a) -- (\b);

\node[inputlabel] (aud)  at (\xA,\yS) {Audio\\10\,s, 16\,kHz};
\node[student]    (vgg)  at (\xB,\yS) {VGGish\,\snow\\$512$-d};
\node[student]    (tr)   at (\xC,\yS) {Transformer\,\fire\\1\,layer, 4\,heads};
\node[student]    (head) at (\xD,\yS) {Binary head\,\fire};
\node[logitnode]  (zs)   at (\xE,\yS) {$\mathbf{z}_s$};
\foreach \a/\b in {aud/vgg, vgg/tr, tr/head, head/zs} \draw[arrsolid] (\a) -- (\b);

\node[logitnode, minimum width=14mm] (y) at (\xY, \yY) {$\mathbf{y}$ (GT)};

\node[masknote] (msk) at (\xRight, \yT) {%
  \textbf{TFD mask}\\[1pt]
  $m{=}0$ if $y{=}\mathrm{Ped}\;\land\;\sigma(\mathbf{z}_t)_0{>}\tau$
};
\node[loss] (lss) at (\xRight, \yS) {%
  $\ell = \ell_{\mathrm{CE}}(\mathbf{z}_s,\mathbf{y})$\\[1pt]
  $+\;\alpha\,m\,\ell_{\mathrm{KD}}(\mathbf{z}_s,\mathbf{z}_t)$
};

\draw[arrdash] (zt) -- (msk.west);
\draw[arrdash] (zs) -- (lss.west);
\draw[arrdash] (y.east) -- ++(0.45,0) |- (msk.west);
\draw[arrdash] (msk.south) -- (lss.north);

\node[phaselab, color=gray!55!black, anchor=west] at (\xA-0.2, \yT+0.62) {Training only (dashed arrows): teacher branch + KD loss};
\node[phaselab, color=green!45!black, anchor=west]   at (\xA-0.2, \yS-0.62) {Training + Inference (solid arrows): audio-only student};

\begin{scope}[on background layer]
  \node[fill=gray!4, rounded corners=4pt, fit=(vid)(m2f)(filt)(mlp)(zt), inner xsep=2.5mm, inner ysep=2.2mm] {};
  \node[fill=green!5,  rounded corners=4pt, fit=(aud)(vgg)(tr)(head)(zs), inner xsep=2.5mm, inner ysep=2.2mm] {};
\end{scope}

\end{tikzpicture}%
}
\vspace{-10pt}
\caption{System pipeline (\fire~trainable, \snow~frozen). Teacher supervision is used only during training, and inference is audio-only. The \textbf{TFD mask} gates the logit and hybrid losses via Eq.~\ref{eq:tfd_label} ($\tau{=}0.4$), and the masked KD term is aggregated by Eq.~\ref{eq:tfd}.}
\label{fig:pipeline}
\vspace{-10pt}
\end{figure*}

Fig.~\ref{fig:pipeline} summarizes the pipeline.

\vspace{3pt}
\textbf{Dataset}: ASPED~v.a~\cite{seshadri2024asped} contains 5 outdoor sessions with synchronized audio and video in urban pedestrian areas closed to vehicular traffic. Ground-truth (GT) labels come from Mask2Former~\cite{cheng2022mask2former} (Swin-S~\cite{liu2021swin}, threshold $0.7$) at $1$~frame per second. A one-second sample is Ped when a detected person lies within $6$\,m of the recorder. The label distribution is $91.5\%$ no-pedestrian (No-Ped) and $8.5\%$ pedestrian (Ped).

\vspace{3pt}
\textbf{Models}: The student follows~\cite{kim2025audio}: a frozen VGGish encoder~\cite{hershey2017vggish} pretrained on AudioSet~\cite{gemmeke2017audioset}, a trainable single-layer 4-head Transformer, and a binary head. Following~\cite{seshadri2024asped, kim2025audio}, every configuration including the baseline counters the imbalance identically, with inverse-frequency batch sampling and a class-weighted binary cross-entropy ($w_c \propto 1/\sqrt{n_c}$, positive weight $3.28$). $\mathcal{L}_{\mathrm{CE}}$ denotes this weighted form throughout. The teacher head is fit on the GT labels, so its logits are an imperfect readout of the GT rather than an independent signal (Section~\ref{sec:teacher_diagnostic}). A frozen Mask2Former (Swin-B, threshold $0.5$) processes a rectangular crop covering the same $6$\,m ground region as the label, extruded to a $2$\,m height. The embedding of its highest-scoring person query ($256$-d) plus confidence and person-count channels ($258$-d) feeds a frozen MLP head ($258{\to}128{\to}2$), whose soft logits $\mathbf{z}_t$ are the distillation target. The head is fit per fold on the four training sessions alone, so no fold's teacher has seen its held-out session, and the student uses no teacher at inference.

\vspace{3pt}
\textbf{Compared configurations}: We compare ten KD configurations against the KD-free baseline (CE loss only) in Table~\ref{tab:final}:
\vspace{2pt}
\begin{compactenum}
\item \emph{Logit-based KD}. LogitKD (Eq.~\ref{eq:kd_loss}, $T{=}3$) and LogitKD$_{0.3}$ with reduced weight $\alpha{=}0.3$.
\vspace{2pt}
\item \emph{Imbalance-aware KD}. LogitKD$_{\mathrm{EMA}}$ (the KD loss normalized by its exponential moving average), LogitKD$_{\mathrm{Adj}}$ (logit adjustment~\cite{menon2021logit}), and LogitKD$_{\mathrm{Focal}}$ (focal weighting~\cite{lin2017focal}).
\vspace{2pt}
\item \emph{Feature-based and relation-based KD}. KD$_{\mathrm{Cos}}$ (cosine similarity between the teacher and student embeddings), KD$_{\mathrm{CRD}}$~\cite{tian2020crd}, and KD$_{\mathrm{RKD}}$~\cite{park2019rkd}.
\end{compactenum}
\vspace{3pt}
The proposed configurations are LogitKD$_{\mathrm{TFD}}$ and Hybrid$_{\mathrm{TFD}}$, where Hybrid adds the cosine term of KD$_{\mathrm{Cos}}$ to the logit loss at the same weight $\alpha$ (Section~\ref{sec:setup}). All share the optimizer, schedule, and student architecture, so only the loss and the per-sample gate vary.

\vspace{3pt}
\textbf{Metrics}: We report \emph{macro accuracy}, the mean of the per-class accuracies $\mathrm{Acc}_{\mathrm{Ped}}$ and $\mathrm{Acc}_{\text{No-Ped}}$ at the $0.5$ threshold, as in~\cite{seshadri2024asped, kim2025audio}. $\mathrm{F1}_{\mathrm{Ped}}$ is the primary minority-class metric, summarizing pedestrian precision and recall without averaging over the No-Ped class, and like macro accuracy it is computed at the $0.5$ threshold. Two threshold-free metrics complement these. \emph{PR-AUC} (Precision-Recall Area Under the Curve, computed as average precision, higher is better) measures how well the student ranks Ped above No-Ped, our probe for transferred discriminative knowledge. \emph{ECE} (Expected Calibration Error, $10$ equal-width bins, lower is better)~\cite{guo2017calibration} is reported both as the standard top-1-confidence ECE (confidence vs.\ accuracy) and as a positive-class ECE that bins the predicted pedestrian probability against its empirical frequency. Since a single score can mix the two effects, we also report the decomposition of the Brier score (BS)~\cite{brier1950, murphy1973}. With $n_k$ of $n$ samples in bin $k$ ($10$ equal-width bins of the predicted Ped probability), mean predicted probability $\bar{p}_k$, positive rate $\bar{y}_k$, and base rate $\bar{y}$,
\begin{equation}
\mathrm{BS}_{\mathrm{binned}} \;=\; \mathrm{REL} - \mathrm{RES} + \bar{y}(1-\bar{y}),
\label{eq:brier}
\end{equation}
where $\mathrm{REL}=\frac{1}{n}\sum_{k} n_k(\bar{p}_k-\bar{y}_k)^{2}$ is the reliability (calibration gap, lower is better), $\mathrm{RES}=\frac{1}{n}\sum_{k} n_k(\bar{y}_k-\bar{y})^{2}$ is the resolution (discrimination, higher is better), and the last term is fixed by the data. This decomposition is exact for the binned forecasts (bin means), a $10$-bin approximation to the continuous Brier score, so we read resolution as corroborating PR-AUC rather than replacing it (Section~\ref{sec:calibration}).

\vspace{3pt}
\textbf{Experiments}: We run two experiments.
\vspace{2pt}
\begin{compactenum}
\item \emph{5-fold Leave-One-Session-Out (LOSO) cross-validation} ($7.7$--$8.4$M labeled seconds per fold across the four training sessions) is our main protocol. Metrics are reported as 5-fold means, with the per-fold std summarized in the caption of Table~\ref{tab:final}. Configurations are compared to the baseline on per-fold macro accuracy with Cohen's $d$ (population standard deviation) as effect size. At five folds an exact two-sided sign-flip test cannot reach $p{<}0.05$ (its floor is $0.0625$), so we prioritize $d$ and per-fold consistency, reporting paired $t$-test $p$-values as supplementary evidence.
\vspace{2pt}
\item The \emph{teacher evaluation} (Section~\ref{sec:teacher_diagnostic}) scores the video teacher on the LOSO folds to contextualize the student results.
\end{compactenum}
\vspace{3pt}

\textbf{Hyperparameters}: Adam optimizer~\cite{kingma2015adam} (lr $5{\times}10^{-4}$, weight decay $0.01$), batch size 64, maximum 2 epochs. Early stopping uses macro accuracy on a random $10\%$ split of the training-session windows. These windows overlap temporally with training, and all reported metrics use the separate held-out session. The default KD weight is $\alpha{=}1.0$. Training windows are drawn at a $1$-second stride ($90\%$ overlap). Each $10$-second window yields ten second-level predictions aligned with the labels, and the held-out session is scored on non-overlapping $10$-second windows, so every test second is predicted exactly once.

%% file: sections/5_results.tex
\vspace{-3pt}
\section{Results and Analysis}
\label{sec:results}
\begin{table}[t]
\centering
\caption{LOSO 5-fold means (accuracies and $\mathrm{F1}_{\mathrm{Ped}}$ in percent, PR-AUC in $[0,1]$, per-fold macro std $1.6$--$2.1$). $d$ is Cohen's $d$ (population std) on per-fold macro differences vs.\ the baseline. Paired $t$-test $p$ by row ($^{*}${:}\,$p{<}0.05$): $0.079$, $0.053$, $0.083$, $0.104$, $0.324$, $0.106$, $0.579$, $0.096$, $0.033^{*}$, $0.026^{*}$. None survives Bonferroni ($0.005$, ten tests).}
\vspace{2pt}
\label{tab:final}
\setlength{\tabcolsep}{4pt}
\footnotesize
\resizebox{\columnwidth}{!}{%
\begin{tabular}{@{}lcccccc@{}}
\toprule
\textbf{Method} & \textbf{Macro}$\uparrow$ & \textbf{No-Ped}$\uparrow$ & \textbf{Ped}$\uparrow$ & $\mathbf{F1_{Ped}}\uparrow$ & \textbf{PR-AUC}$\uparrow$ & $\boldsymbol{d}\uparrow$ \\
\midrule
Baseline (CE)          & $72.9$ & $63.5$ & $82.3$ & $28.6$ & $0.328$ & --      \\
\midrule
\multicolumn{7}{l}{\textit{Logit-based KD}} \\
LogitKD                     & $73.6$ & $70.8$ & $76.4$ & $31.2$ & $0.332$ & $1.17$  \\
LogitKD$_{0.3}$             & $73.8$ & $70.3$ & $77.2$ & $30.9$ & $0.326$ & $1.36$  \\
\midrule
\multicolumn{7}{l}{\textit{Imbalance-aware KD}} \\
LogitKD$_{\mathrm{EMA}}$    & $73.6$ & $68.5$ & $78.7$ & $30.4$ & $0.324$ & $1.15$  \\
LogitKD$_{\mathrm{Adj}}$    & $71.3$ & $53.7$ & $89.0$ & $25.9$ & $0.333$ & $-1.05$ \\
LogitKD$_{\mathrm{Focal}}$  & $73.3$ & $65.2$ & $81.4$ & $29.3$ & $0.329$ & $0.56$  \\
\midrule
\multicolumn{7}{l}{\textit{Feature- and relation-based KD}} \\
KD$_{\mathrm{Cos}}$         & $73.6$ & $67.4$ & $79.8$ & $30.1$ & $0.327$ & $1.04$  \\
KD$_{\mathrm{CRD}}$         & $73.2$ & $64.4$ & $82.0$ & $29.0$ & $0.329$ & $0.30$  \\
KD$_{\mathrm{RKD}}$         & $73.7$ & $67.3$ & $80.1$ & $30.0$ & $0.327$ & $1.08$  \\
\midrule\midrule
\multicolumn{7}{l}{\textit{Proposed (Trust-Filtered Distillation)}} \\
\textbf{LogitKD$_{\mathrm{TFD}}$} & $73.7$ & $72.6$ & $74.7$ & $31.8$ & $0.330$ & $1.60^{*}$          \\
\textbf{Hybrid$_{\mathrm{TFD}}$}  & $73.6$ & $72.9$ & $74.3$ & $31.8$ & $0.331$ & $1.72^{*}$ \\
\bottomrule
\end{tabular}}
\vspace{-12pt}
\end{table}

Table~\ref{tab:final} reports the main cross-session comparison. Two observations frame it. First, no configuration gains more than one point of macro accuracy and the fold-to-fold std exceeds these gains, so we compare fold by fold with paired tests rather than by means. Second, $\mathrm{Acc}_{\mathrm{Ped}}$ (Ped recall) and $\mathrm{F1}_{\mathrm{Ped}}$ diverge throughout ($82.3$ vs.\ $28.6$ for the baseline), because under $91.5\%$ imbalance even a moderate false-positive rate outnumbers the pedestrians and keeps precision, hence F1, low. Our claims therefore concern relative differences and mechanisms rather than absolute performance.

\subsection{Cross-Session Method Comparison (Experiment 1)}
\label{sec:loso}

PR-AUC ranges from $0.324$ to $0.333$ across configurations (baseline $0.328$). Among configurations with per-fold predictions, the largest mean PR-AUC gain is $+0.004$ for LogitKD (descriptive $95\%$ bootstrap CI $[0.002,0.008]$), and the corresponding intervals include zero for LogitKD$_{\mathrm{TFD}}$ $[-0.002,0.006]$ and Hybrid$_{\mathrm{TFD}}$ $[-0.002,0.008]$.

Comparison at a matched operating point indicates only small discrimination gains. Moving the baseline along each fold's ROC to LogitKD's false-positive rate reproduces most of the recall change, leaving only about $+1.5$ recall ($74.9$ vs.\ $76.4$) and $+0.4$ F1 ($30.8$ vs.\ $31.2$). ROC-AUC rises from $0.808$ to $0.813$--$0.815$, and LogitKD$_{\mathrm{TFD}}$ shows no consistent matched-FPR gain over LogitKD (three models with saved predictions).

The two TFD configurations have the largest standardized paired macro-accuracy effects against the baseline (Table~\ref{tab:final}). Directly comparing LogitKD with and without the gate, however, reveals no significant difference across five folds in macro accuracy ($p{=}0.84$) or $\mathrm{F1}_{\mathrm{Ped}}$ ($p{=}0.14$). The gate also renormalizes the surviving KD term (Section~\ref{sec:tfd}), so this comparison does not isolate sample selection from the effective KD weight, and TFD accentuates the operating-point shift without establishing an advantage over ungated KD.

\vspace{-4pt}
\subsection{Per-Class Operating-Point Shift (Experiment 1)}
\label{sec:opshift}

KD raises $\mathrm{Acc}_{\text{No-Ped}}$ and lowers $\mathrm{Acc}_{\mathrm{Ped}}$ (LogitKD: $63.5{\to}70.8$ and $82.3{\to}76.4$), and the direction holds for every configuration except LogitKD$_{\mathrm{Adj}}$, which over-corrects toward Ped. This class-wise change is an asymmetric operating-point shift that trades pedestrian recall for majority-class accuracy, consistent with the teacher-target asymmetry in Section~\ref{sec:setup}. For the three models with per-fold predictions, the baseline's macro accuracy is nearly flat across this operating range ($\approx\!72.9$), so the small macro gain ($+0.7$) is not reproduced by thresholding the baseline (a small, single-seed effect). $\mathrm{F1}_{\mathrm{Ped}}$ rises ($28.6$ to $31.8$ for LogitKD$_{\mathrm{TFD}}$), but the baseline reaches $31.4$ at the same false-positive rate, so most of this rise is an operating-point effect rather than better minority-class detection. The lower recall is thus a changed precision--recall trade-off, not a uniform minority-class loss.

\vspace{-4pt}
\subsection{Calibration vs.\ Probability Ranking (Experiment 1)}
\label{sec:calibration}

PR-AUC changes only slightly (Section~\ref{sec:loso}). Calibration metrics that bin the predicted pedestrian probability improve (positive-class ECE $0.366{\to}0.316$, Brier reliability $0.163{\to}0.110$), but reduced over-prediction dominates them. The positive-class ECE closely tracks the mean prediction's distance from the per-fold prior ($0.366$, $0.316$), and by Jensen's inequality $(\bar{p}-\pi)^2$ lower-bounds the reliability, equaling $83\%$ and $91\%$ of the fold-mean reliability. The mean predicted probability falls from $0.45$ to $0.40$ while resolution is unchanged at $0.011$. The standard top-1-confidence ECE is nearly unchanged ($0.076$ vs.\ $0.073$) but masks offsetting per-class miscalibration, as Ped predictions are overconfident (about $18\%$ accurate) while No-Ped predictions are underconfident and binning by confidence cancels the two. The aggregate calibration gains thus reflect reduced over-prediction and an operating-point shift rather than better ranking.

\vspace{-4pt}
\subsection{Teacher Comparison (Experiment 2)}
\label{sec:analysis}\label{sec:teacher_diagnostic}

The sampler balances the input classes in expectation, so the shift is not one of batch prevalence. It is consistent instead with the teacher's lower mean confidence on Ped (Section~\ref{sec:setup}), and the student's predicted Ped ratio falls from $40.4\%$ to $33.2\%$ under KD. Because TFD removes KD only on Ped samples, both the KD-supervised set and its effective weight (via the survivor renormalization, Section~\ref{sec:tfd}) move toward No-Ped, consistent with TFD's stronger shift, though no random-mask control was run. The gate targets the observed $91{:}9$ imbalance, and neither its optimality at other class ratios nor its causal role in the shift is established.

The video teacher reaches $89.9\%$ LOSO macro accuracy ($\mathrm{F1}_{\mathrm{Ped}}$ $53\%$) against the student's $72.9\%$ and $28.6\%$, yet the student's PR-AUC changes only slightly across the evaluated KD configurations (Section~\ref{sec:loso}), so the teacher's higher task performance does not translate into comparably large gains in student ranking. This is consistent with the Modality Focusing Hypothesis~\cite{xue2023modality}, but our single pipeline does not isolate the modality gap from teacher calibration, student capacity, or optimization effects.

%% file: sections/7_conclusion.tex
\vspace{-5pt}
\section{Conclusion}
\label{sec:conclusion}

Across ten KD configurations on ASPED, modest macro-accuracy gains generally accompany an operating-point shift toward the majority class, with only small changes in PR-AUC. We analyze Trust-Filtered Distillation, a ground-truth-conditioned, minority-selective gate whose logit formulation admits a conditional label-smoothing interpretation under a shared temperature. Our evidence is limited to ASPED (one campus, five folds, binary, single runs, uncorrected for multiple comparisons), and it does not isolate the gate's selectivity from its reweighting or test other imbalance severities, which future work should address alongside closing the modality gap. Code is released at \url{https://github.com/yonghyunk1m/cmkd-ped-detect}.

\vspace{-5pt}
\section{Acknowledgment}
The authors used Anthropic's Claude and Microsoft Copilot for manuscript editing. All research ideas, experiments, results, and analysis are the authors' own, who take full responsibility.